\documentclass[journal=jpclcd,manuscript=article,layout=twocolumn]{achemso}
\setkeys{acs}{articletitle = true}

\usepackage{graphicx}
\usepackage{dcolumn}
\usepackage{amsmath}
\usepackage{amssymb}
\usepackage{subfigure,amsmath,verbatim,moreverb,bm}
\usepackage{booktabs}
\usepackage{nicefrac}
\usepackage{natbib}

\usepackage{dcolumn}   

\def\beq{\begin{equation}}
\def\eeq{\end{equation}}
\def\ber{\begin{eqnarray}}
\def\eer{\end{eqnarray}}

\def\rv{{\bf r}}

\def\vw{{\mathbf W}}

\usepackage{color}

\author{Stefan Vuckovic}
\email{s.vuckovic@vu.nl}
\affiliation{Department of Theoretical Chemistry and Amsterdam Center for Multiscale Modeling, FEW, Vrije Universiteit, De Boelelaan 1083, 1081HV Amsterdam, The Netherlands}
\author{Paola Gori-Giorgi}
\affiliation{Department of Theoretical Chemistry and Amsterdam Center for Multiscale Modeling, FEW, Vrije Universiteit, De Boelelaan 1083, 1081HV Amsterdam, The Netherlands}
\author{Fabio Della Sala}
\affiliation{Institute for Microelectronics and Microsystems (CNR-IMM), Via Monteroni, Campus Unisalento, 73100 Lecce, Italy}
\affiliation{Institute for Microelectronics and Microsystems (CNR-IMM), Via Monteroni, Campus Unisalento, 73100 Lecce, Italy}
\author{Eduardo Fabiano}
\affiliation{Institute for Microelectronics and Microsystems (CNR-IMM), Via Monteroni, Campus Unisalento, 73100 Lecce, Italy}
\affiliation{Institute for Microelectronics and Microsystems (CNR-IMM), Via Monteroni, Campus Unisalento, 73100 Lecce, Italy}

\title{Restoring Size Consistency of Approximate Functionals Constructed from the Adiabatic Connection}

\begin{document}
\begin{tocentry}
\includegraphics{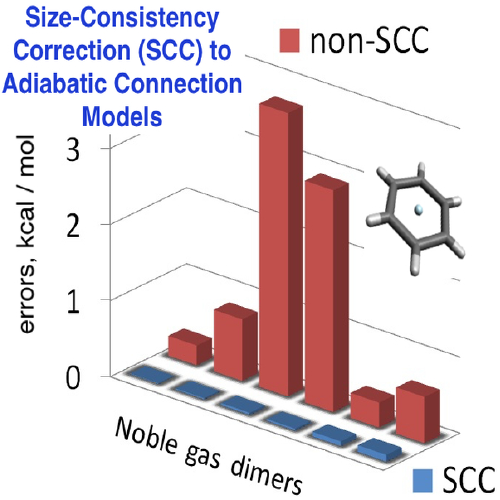}
\end{tocentry}
\begin{abstract}
Approximate exchange-correlation functionals built by modeling in a
  non-linear way the adiabatic connection (AC) integrand of density functional
  theory have many attractive features, being virtually parameters-free and 
 satisfying different exact properties, but they also have a
  fundamental flaw: they violate the size-consistency condition, crucial to
  evaluate interaction energies of molecular systems. 
  We show that size consistency in the AC-based functionals can be restored in a
  very simple way at no extra computational cost. 
  Results on a large set of benchmark molecular interaction energies show that
  functionals based on the interaction strength interpolation approximations are
  significantly more accurate than the second-order perturbation theory.
\end{abstract}


With applications that stretch from solid state physics to biochemistry,
Kohn-Sham density functional theory (KS-DFT)\cite{KohSha-PR-65} is presently
the most employed electronic structure method. Although the theory is in
principle exact, any practical implementation of KS-DFT must rely on
approximations for the exchange-correlation (XC) functional, which should
capture all the many-body effects beyond the simple Hartree theory.
 Despite the existence of hundreds of different XC approximations
 \cite{MarHea-MP-17,metarev} and their widespread success in various
 disciplines \cite{MarHea-MP-17}, KS-DFT still encounters open issues, which
 hamper its overall predictive power
 \cite{CohMorYan-CR-12,Bur-JCP-12,Bec-JCP-14,MarHea-MP-17} and make the quest
 for better approximations a crucial research field for computational
 chemistry, solid state physics and materials
 science.\cite{CohMorYan-CR-12,Bur-JCP-12,Bec-JCP-14,MarHea-MP-17}

The density-fixed adiabatic connection (AC) formalism
\cite{LanPer-SSC-75,GunLun-PRB-76} provides an exact expression for the XC
energy functional $E_{xc}[\rho]$,
\beq \label{eq:acxc}
E_{xc}[\rho]= \int_0^1 W_\lambda[\rho]\mathrm{d}\lambda,
\eeq
where $W_\lambda[\rho]$ is the AC integrand,
\beq \label{eq:Wlam}
W_\lambda[\rho]= \langle \Psi_{\lambda}[\rho]  | \hat{V}_{ee}| \Psi_{\lambda}[\rho]\rangle - U[\rho],
\end{equation}
$\Psi_{\lambda}[\rho]$ is the fermionic wavefunction with density
$\rho(\rv)$ that minimizes the sum of the kinetic energy $\hat{T}$ and of the
electron-electron repulsion $\hat{V}_{ee}$ scaled by the coupling constant
$\lambda$, and $U[\rho]$ is the Hartree energy.
For small systems, $W_\lambda[\rho]$ has also been
computed exactly through eq~\ref{eq:Wlam}.
However, this requires  the solution of 
the many-body Schr\"{o}dinger equation.\cite{ColSav-JCP-99,TeaCorHel-JCP-09,TeaCorHel-JCP-10}
Thus, to all practical purposes $W_\lambda[\rho]$ must be approximated.

Equation~\ref{eq:acxc} has been a fundamental milestone in 
guiding the construction of approximations. 
Early AC-based XC functionals used forms that depend linearly on some
chosen input ingredients, such as the exchange energy from Hartree-Fock theory
as value to be recovered at $\lambda=0$, and semilocal approximations at some
$\lambda=\lambda_p$ between 0 and 1. These forms are commonly used for the
construction of hybrid \cite{Bec-JCP-93a,Bec-JCP-93,PerErnBur-JCP-96} and
double-hybrid \cite{Gri-JCP-06,LarGri-JCTC-10,ShaTouSav-JCP-11} density
functionals, resulting in mixing a fixed fraction of Hartree-Fock exchange and
second-order perturbation theory with semilocal functionals. They often work
well for main-group chemistry, but they show important limitations for various
other problems as for example the chemistry of transition metals\cite{CraTru-PCCP-09}
(where they even worsen the results with respect to simpler
semilocal functionals), metal-molecule interfaces
\cite{fabiano09}, and even non-covalent bonding (unless an ad hoc van der Waals
correction is used).\cite{corminboeuf14,hapbe}
Their main disadvantage is that the
mixing fractions are fixed and cannot adapt to different systems or to
different parts of a system.

To address this problem, several models in which the input ingredients enter
in a non-linear way have been
proposed.\cite{Ern-CPL-96,SeiPerLev-PRA-99,SeiPerKur-PRL-00,MorCohYan-JCP-06,TeaCorHel-JCP-10,locpaper}
These latter forms do not need to rely on empiricism, and can adapt automatically to
the peculiarities of the system under study. Along these lines, Ernzerhof had
proposed Pad{\'e} forms for the $\lambda$-dependence of the AC
integrand\cite{Ern-CPL-96}, which later were used for the construction of the
MCY family of functionals that are constrained to be free of one-electron
self-interaction error.\cite{MorCohYan-JCP-06,CohMorYan-JCP-07} Another
example of models that use input ingredients in a non-linear way is provided
by the interaction strength interpolation (ISI) functionals, which depend
explicitly on the weak- and strong-coupling ingredients
\cite{SeiPerLev-PRA-99,SeiPerKur-PRL-00,GorVigSei-JCTC-09,LiuBur-PRA-09,JiaEng-ZPC-10},
essentially extending to non-uniform densities Wigner's
\cite{Wig-PR-34,Wig-TFS-38} idea for approximating the energy of the
homogeneous electron gas. Despite the advantages of the nonlinear forms over
the linear ones, the former encounter a fundamental flaw: the XC functionals
that are constructed from them are not size-consistent for systems composed by
different species of fragments.\cite{CohMorYan-JCP-07,MirSeiGor-JCTC-12}
This depends on the fact that these methods employ as input ingredients global quantities (integrated over all space). 
A route that is currently being explored addresses this issue by modeling the
AC at each given spatial position $\rv$, using energy densities
$w_\lambda(\rv)$
\cite{locpaper,ZhoBahErn-JCP-15,BahZhoErn-JCP-16,VucIroWagTeaGor-PCCP-17}
instead of quantities integrated over all space. This strategy is very
promising, but does not allow using in a straightforward way semilocal
ingredients
\cite{JarScuErn-JCP-03,MaiHasArbKau-PCCP-16,MirSeiGor-JCTC-12,locpaper,ZhoBahErn-JCP-15,BahZhoErn-JCP-16,VucIroWagTeaGor-PCCP-17,VucLevGor-JCP-17},
because of the inherent ambiguity in the definition of energy densities, a
problem shared with the construction of local hybrid functionals.
\cite{JarScuErn-JCP-03,ArbKau-CPL-07,BahKau-JCTC-15}

In this Letter we show that size consistency of the {\em global} (integrated
over all space) AC forms in which the ingredients enter in a non-linear way
can be restored in a remarkably simple way, making it possible to obtain
meaningful interaction energies at no additional computational cost.

Consider a system $M$ (e.g., a 
molecular complex) composed of a set of fragments $A_i$, 
with $i=1,\ldots,N$. The interaction energy is a key quantity in chemistry and it is defined as
\beq \label{eq:intertot}
E^{\rm int}(M)=E(M)-\sum_{i=1}^N E(A_i),
\eeq
where $E(M)$ is the energy of the bound system $M$ and $E(A_i)$ are the
energies of the individual fragments. If we now compute the energy of a system
$M^*$ made of the same fragments $A_i$ placed at a very large (infinite)
distance from each other, any size-consistent method should give $E^{\rm
  int}(M^*)=0$, or equivalently

\beq \label{eq:sizeconsistency}
E(M^*)=\sum_{i=1}^N E(A_i).
\eeq

 We should stress at this point that size consistency in DFT is in general a very subtle issue, particularly 
when dealing with fragments with a degenerate ground state (e.g., open-shell atoms), as the (spin) density is not anymore an intensive quantity.\cite{GorSav-JPCS-08,Sav-CP-09} To disentangle this more general DFT problem from the one of size-consistency of the non-linear AC models, here we focus on the cases where the fragments have a non-degenerate
  ground-state \cite{notedeg}, considering non-covalent interactions.

The idea behind AC-based functionals is to use a certain number of input
ingredients $W_i[\rho]$, constructing a $\lambda$-dependent function that
interpolates between them.  For example, many standard hybrid functionals
model $W_\lambda[\rho]$ with functions of the kind
\begin{equation}
	\label{eq:W_lambdaHYB}
W_\lambda^{\rm hyb}[\rho]=W_\lambda^{\rm DFA}[\rho]+(E_x^{\rm HF}[\rho]-E_x^{\rm DFA}[\rho])\lambda^{n-1},
\end{equation}
where $W_\lambda^{\rm DFA}[\rho]$ is a given density-functional approximation
(usually a semilocal functional), with its exchange component
$W_{\lambda=0}^{\rm DFA}[\rho]=E_x^{\rm DFA}[\rho]$, and $E_x^{\rm HF}[\rho]$
is the Hartree-Fock (HF) exchange energy. 

This kind of expressions, when inserted in eq~\ref{eq:acxc}, yields a fixed
fraction $1/n$ of HF exchange energy mixed with a semilocal density functional
approximation. Because the input ingredients, in this case $W_\lambda^{\rm
  DFA}[\rho]$, $E_x^{\rm HF}[\rho]$, and $E_x^{\rm DFA}[\rho]$, enter linearly
in the model of eq~\ref{eq:W_lambdaHYB}, the resulting XC functional
automatically satisfies the size-consistency condition of
eq~\ref{eq:sizeconsistency} if the individual ingredients do.

As examples of approximations in which the ingredients enter in a non-linear
way, consider first the Pad{\'e}([1/1]) form introduced by Ernzerhof
\cite{Ern-CPL-96},
\begin{equation}
	\label{eq:Pade}
	W_\lambda^{\rm Pad}[\rho]=a[\rho]+\frac{b[\rho]\,\lambda}{1+c[\rho]\,\lambda},
\end{equation}
with $a[\rho]=W_0[\rho]=E_x^{\rm HF}[\rho]$, $b[\rho]=W_0'[\rho]$ (which can
be obtained from second-order perturbation theory), and
$c[\rho]=\lambda_p^{-1}-W_0'[\rho](W_0[\rho]-W_{\lambda_p}^{\rm
  DFA}[\rho])^{-1}$, where  $W_{\lambda_p}^{\rm DFA}[\rho]$ could be a
semilocal functional at a chosen value $\lambda_p$. We see immediately that in
this case, even if the input quantities $W_0$, $W_0'$ and $W_{\lambda_p}$
satisfy the size-consistency condition of eq~\ref{eq:sizeconsistency}, the
resulting XC functional from eq~\ref{eq:acxc} does not, because it is given
by a non-linear function $f^{\rm Pad}$ of these ingredients, 
$E_{xc}^{\rm Pad}=f^{\rm Pad}(W_0,W_0',W_{\rm \lambda_p})$.

Another example is the idea of Seidl and co-workers
\cite{SeiPerLev-PRA-99,SeiPerKur-PRL-00} to build approximate
$W_\lambda[\rho]$ by interpolating between its weak- ($\lambda\to 0$) and
strong- ($\lambda\to\infty$) coupling expansions,

\begin{eqnarray}
\label{eq:lambda0}
W_{\lambda\rightarrow0}[\rho] & = & W_0[\rho] + \lambda W_0'[\rho] + \cdots \\
\label{eq:lambdainf}
W_{\lambda\rightarrow\infty}[\rho] & = & W_{\infty}[\rho] + \frac{W_{\infty}'[\rho]}{\sqrt{\lambda}}  + \cdots\ ,
\end{eqnarray}
which allows avoiding bias towards the weakly correlated regime, and to
include more pieces of exact information. The $\lambda\to 0$ limit of
eq~\ref{eq:lambda0} is provided by the exact exchange and the second-order
perturbation theory, while the functionals $W_{\infty}[\rho]$ and
$W_{\infty}'[\rho]$ describe a floating Wigner crystal with a metric dictated
by the density.\cite{SeiGorSav-PRA-07,GorVigSei-JCTC-09}

Different formulas that interpolate between the limits of
eqs~\ref{eq:lambda0} and \ref{eq:lambdainf} are available in the
literature
.\cite{SeiPerLev-PRA-99,SeiPerKur-PRL-00,GorVigSei-JCTC-09,LiuBur-JCP-09,locpaper} As
in the Pad\'e example of eq~\ref{eq:Pade}, when these interpolation
formulas are inserted in eq~\ref{eq:acxc} they give an XC energy that is a
nonlinear function of the 4 ingredients (or a subset thereof)
$W_0[\rho],W'_0[\rho],W'_\infty[\rho],W_\infty[\rho]$  appearing in
eqs~\ref{eq:lambda0}-\ref{eq:lambdainf}. 

It is clear from these examples that we can write a general XC functional obtained by modeling the adiabatic connection as 
\begin{equation}
E_{xc}^{\rm ACM}[\rho]=f^{\rm ACM}(\vw[\rho]), 
\end{equation}
where $f^{\rm ACM}$ is a non-linear function that results from the integration
[via eq\ref{eq:acxc}] of the given adiabatic connection model (ACM), and
$\mathbf{W}[\rho]=\{W_1[\rho],...,W_k[\rho]\}$ 
is a compact notation for the $k$ input ingredients that have been used. Then we have
\beq \label{eq:SumAiACM}
\sum_{i=1}^N E_{xc}^{\rm ACM}(A_i)=\sum_{i=1}^N f^{\rm ACM} \left(  \mathbf{W}(A_i) \right),
\eeq
and 
\beq \label{eq:mstarACM}
E_{xc}^{\rm ACM}(M^*)=f^{\rm{ACM}} \left(\sum_{i=1}^N\mathbf{W}(A_i) \right).
\eeq
This equation is one of the main points in this work. Although conceptually simple, it shows that the energy of a set of infinitely distant fragments ($M^*$) can be expressed as a function of the quantities of the isolated fragments. Notice that this holds in this special case because $f^{ACM}$ is a function of global size-consistent\cite{notedeg} quantities. For size-inconsistent wavefunction methods, such as CISD, this is usually not true, and the energy of $M^*$ needs to be computed by performing an extra calculation with the fragments at a large distance, which might be tricky to do in practice. 

Essentially all the models that have been proposed in the literature \cite{CohMorYan-JCP-07} satisfy the condition 
\begin{equation}
	\label{eq:sizeextens}
f^{\rm ACM} \left(  N\mathbf{W}(A) \right)=Nf^{\rm ACM} \left(  \mathbf{W}(A) \right),
\end{equation}
meaning that they are size-consistent when a system dissociates into equal
fragments (size-extensivity). This is also a key difference with the size-consistency problem of wavefunction methods, which also arises in the case of equal fragments. However, when the $A_i$ are of different
species, eqs~\ref{eq:SumAiACM} and \ref{eq:mstarACM} give in general
different results, and attempts to make them equal for a non-linear model have
failed so far.\cite{CohMorYan-JCP-07}

As said, evaluating
eq~\ref{eq:SumAiACM} or eq~\ref{eq:mstarACM} has exactly the same
computational cost, as both equations only need the input ingredients for the
individual fragments. The idea behind the size-consistency correction (SCC) is
thus extremely simple and it is related to discussions reported in
Refs.~\cite{Hob-CR-88,FabGorSeiSal-JCTC-16,smiga17}: it consists in
using the difference between eq~\ref{eq:mstarACM} and
eq~\ref{eq:SumAiACM} to cancel the size-consistency error that is made when
evaluating interaction energies from eq~\ref{eq:intertot},

\begin{equation}
	\label{eq:DeltaSCC}
	\Delta_{\rm SCC}=\sum_{i=1}^N f^{\rm ACM} \left(  \mathbf{W}(A_i) \right)-f^{\rm{ACM}} \left(\sum_{i=1}^N\mathbf{W}(A_i). \right).
\end{equation}
Note that this correction is fundamentally different from a direct calculation of $\sum_i E(A_i) - E(M^*)$, since, due to the use of eq~(\ref{eq:mstarACM}), only the knowledge of the isolated fragments is required here, while there is no need to deal with the (possibly tricky) calculation of the supramolecular energy $M^*$.

Adding $\Delta_{\rm SCC}$ to an interaction energy computed via eq~\ref{eq:intertot} is equivalent to
always evaluating interaction energies with respect to eq~\ref{eq:mstarACM} instead of eq~\ref{eq:SumAiACM}, i.e.,
\begin{equation}
	\label{eq:SCC}
E_{xc,{\rm int}}^{\rm ACM,SCC}[\rho]=f^{\rm ACM}\left(\vw(M)\right)-f^{\rm{ACM}} \left(\sum_{i=1}^N\mathbf{W}(A_i) \right).
\end{equation}
 \begin{figure}
 \includegraphics[width=1.0\columnwidth]{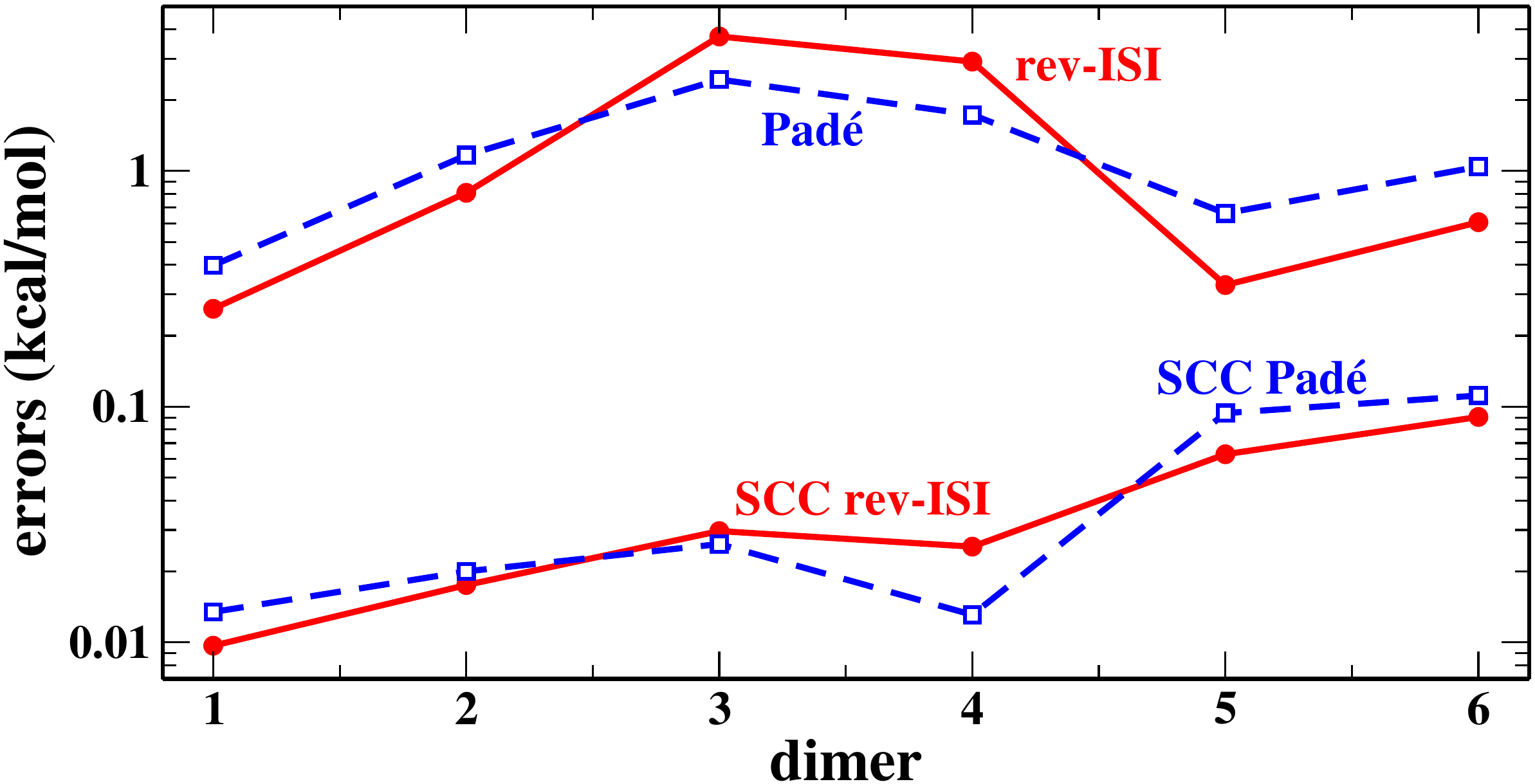}
 \caption{The absolute errors on the interaction energy (kcal/mol, log-scale) for a set
   of dispersion heterodimers, containing the noble gas atoms, obtained with
   the rev-ISI and Pad{\'e}([1/1]) functionals with and without the inclusion
   of the SCC of eq~\ref{eq:DeltaSCC} (1:~He-Ne, 2:~He-Ar, 3:~Ne-Ar,
   4:~Ar-Kr, 5:~CH$_4$-Ne, 6:~C$_6$H$_6$-Ne).}
 \label{fig_het}
 \end{figure}

 \begin{figure}
 \includegraphics[width=1.0\columnwidth]{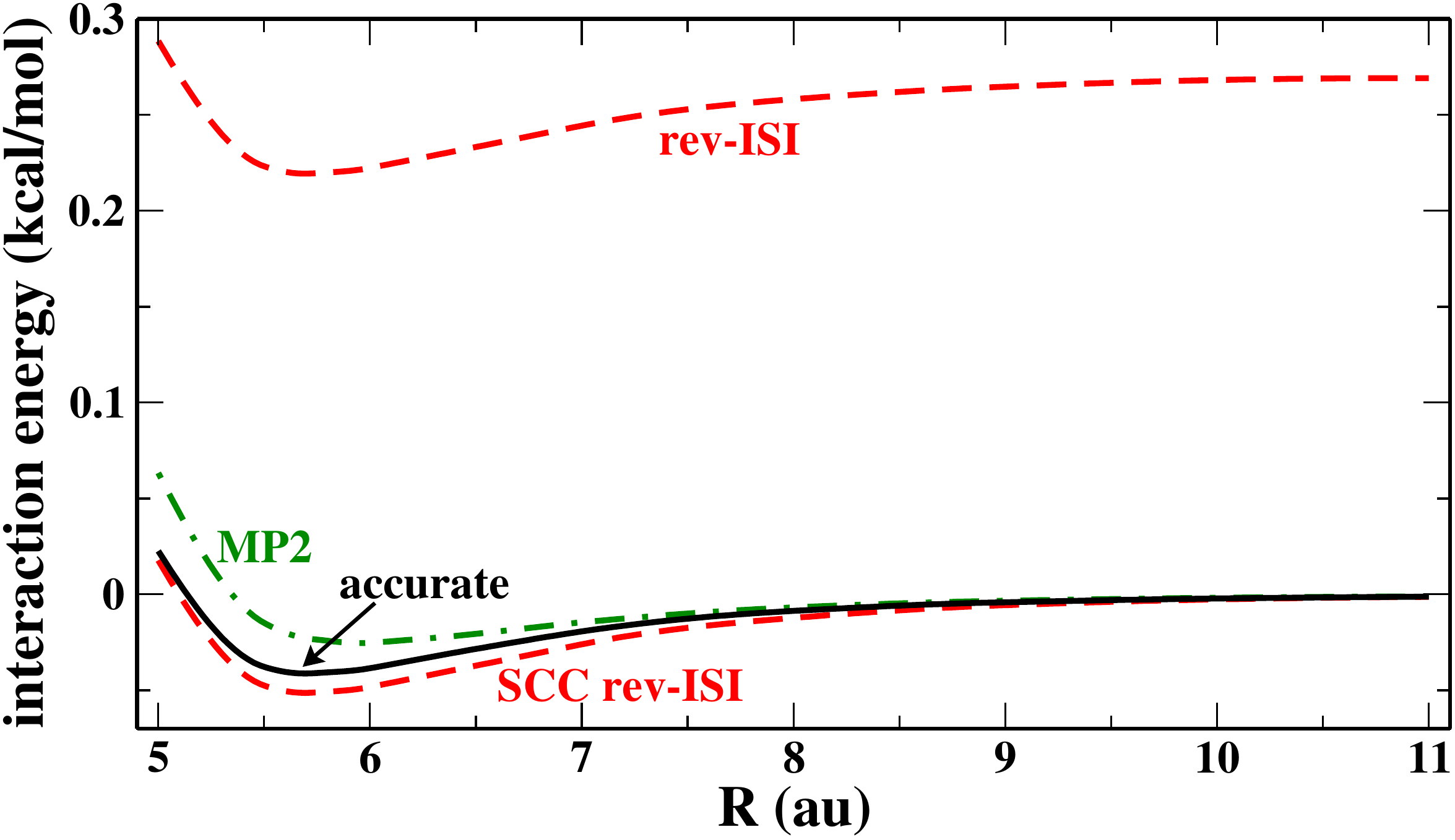}
 \caption{Interaction energy curve for the HeNe heterodimer obtained with the
   rev-ISI functional with and without the SCC of eq~\ref{eq:DeltaSCC}. The
   MP2 curve is shown for comparison and the accurate curve has been taken
   from Ref.~\cite{OgiWan-JMS-93}.}

 \label{fig_hene}
 \end{figure}

As an example of the
performance of the SCC, we examine here ACMs that link the two
limits of eqs~\ref{eq:lambda0} and \ref{eq:lambdainf}. As said, we focus
on non-covalent interactions because the fragments $A_i$ have a non-degenerate
ground state, which should guarantee size-consistency of the input ingredients
.\cite{GorSav-JPCS-08,Sav-CP-09} Moreover, in this case the interaction energy
is small and so the correction can be relevant: for covalent interactions, in
fact, the correction is of the same order of magnitude as for non-covalent
ones, but the interaction energy is at least two orders of magnitude
larger. All calculations have been performed using a development version of
the TURBOMOLE package\cite{turbo,turbo_rev}, with
computational details similar to those of
Refs.~\citenum{FabGorSeiSal-JCTC-16}
and~\cite{GiaGorDelFab-JCP-18}, in which the ISI-like functionals are
evaluated on Hartree-Fock orbitals (see Section IV.D of Ref.~\citenum{GiaGorDelFab-JCP-18} for some discussion of this choice). Thus, in eq~\ref{eq:lambda0}
$W_0[\rho]=E_x^{\rm HF}[\rho]$ and $W_0'[\rho]$ is twice the second-order
M\"oller-Plesset (MP2) correlation energy, $W_0'[\rho]=2 E_c^{\rm MP2}$. The
strong-coupling functionals $W_\infty[\rho]$ and $W_\infty'[\rho]$ of
eq~\ref{eq:lambdainf} are approximated with the point-charge-plus-continuum
(PC) semilocal model \cite{SeiPerKur-PRA-00}, which is reasonably accurate
.\cite{SeiGorSav-PRA-07,GorVigSei-JCTC-09} We test different interpolation
formulas that have been proposed in the literature, namely
SPL\cite{SeiPerLev-PRA-99}, rev-ISI\cite{GorVigSei-JCTC-09} and
LB\cite{LiuBur-JCP-09} 
Additionally,
we also tested the Pad\'e[1,1] formula of eq~\ref{eq:Pade} by using
$\lambda_p=\infty$. The 
 interpolation formulas and additional computational details are reported in the Supporting Information.

As a first example, in Fig.~\ref{fig_het} we show the absolute errors in the
interaction energy for a set of dispersion complexes made of fragments of
different species obtained from the rev-ISI and the Pad\'e interpolation
formulas, computed with and without the SCC. From this figure we can see that
in both cases the error is reduced by an order of magnitude when the
correction is applied, i.e, when eq~\ref{eq:SCC} is used. 

In  Figure~\ref{fig_hene} we also report the interaction energy curve for He-Ne
obtained from the rev-ISI functional. We see that the rev-ISI curve has a very
reasonable shape, but, because of the size-consistency error, when computed
with eq~\ref{eq:intertot} it goes to a positive value with respect to the
sum of the fragment energies. Instead, when the SCC is applied, the
  correct asymptotic value of the dissociation curve (given by
    eq~\ref{eq:mstarACM}) is used to compute interaction energies. Very
similar figures are obtained when we consider other interpolation formulas and
other systems, with the overall shift that is sometimes positive and sometimes
negative. 


Finally, we use the SCC to assess the accuracy of AC-based functionals for
more non-covalent complexes relevant for chemistry and biology. 
For this purpose, we employ the well established quantum-chemical dataset for non-covalent interactions
S66. \cite{RezRilHob-JCTC-11}

\begin{figure}
\includegraphics[width=0.98\columnwidth]{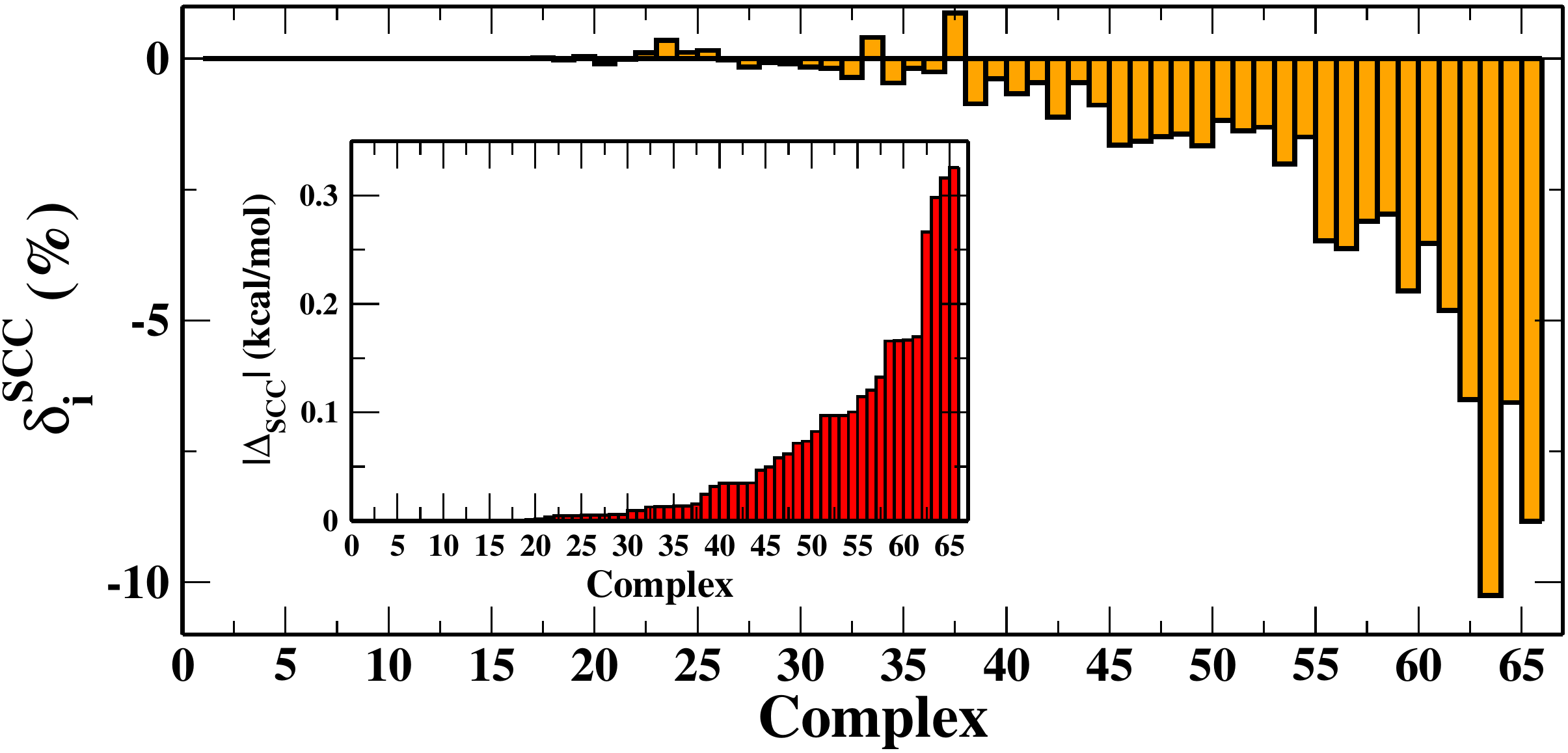}
\caption{\label{fig_s66}Difference $\delta_i^\mathrm{SCC}$ of the absolute
 relative errors on interaction energies calculated with and without
 SCC for the complexes of the S66 test set (sorted with increasing $|\Delta_\mathrm{SCC}|$, see inset).}
\end{figure}

In Figure \ref{fig_s66} we report the values of
\begin{equation}
\delta_i^\mathrm{SCC} = \left|\frac{E_i^\mathrm{rev-ISI-SCC}-E_i^{ref}}{E_i^{ref}}\right|-\left|\frac{E_i^\mathrm{rev-ISI}-E_i^{ref}}{E_i^{ref}}\right|\ ,
\end{equation}
where the index $i=1,\ldots,66$ labels the various complexes, $E_i^{ref}$ is
the reference 
interaction energy of the $i$-th complex and
$E_i^\mathrm{rev-ISI-SCC}$ and $E_i^\mathrm{rev-ISI}$ are the corresponding
interaction energies 
calculated with rev-ISI with and without the SCC 
correction, respectively. Thus a negative (positive) $\delta_i^\mathrm{SCC}$ means 
that the SCC reduces (increases) the relative error of the interaction energies. 
Notice that in  Figure~\ref{fig_s66} the S66 complexes are sorted in ascending order according to the
computed $|\Delta_{\rm SCC}|$ value (see inset of Figure~\ref{fig_s66}).
Thus, one can see that, for systems where $|\Delta_{\rm SCC}|$ is non-negligible (i.e. $i\gtrsim$30), 
the inclusion
of the SCC brings an improvement of the results ($\delta_i^\mathrm{SCC}<0$)
and that the improvement can be as large as 10\%.
On the other hand, there are some systems (i.e. $i\lesssim$30) with a negligible SCC. 
This is not surprising, as the S66 dataset contains 17 homodimers, for which
the AC-based functionals are already size consistent. Moreover, there is
another case in which the size-consistency error becomes negligible: when the
ratio $q_i=W_i(A)/W_i(B)$ between the $i^{th}$ input ingredient of fragment
$A$ and of fragment $B$ is roughly the same for all $i$, $q_i\approx q$, a
case that becomes mathematically equivalent to eq~\ref{eq:sizeextens}.
In summary, Figure~\ref{fig_s66} shows  that 
the larger is the
$\Delta_{\rm SCC}$ value, the larger is the reduction of the errors.
This indicates that the inclusion of $\Delta_{\rm SCC}$  is
significant and works correctly
for most non-covalent complexes having different constituting 
units.

\begin{table*}  
\centering
\caption{\label{tab_s66maes} Mean absolute error (and variance, last column), in kcal/mol, for the
  S66 data set 
and some of its subsets, for different AC-based functionals including SCC 
and evaluated on Hartree-Fock density an orbitals.
{For all results see Table S2 in the Supporting Information.}
The last four lines report, for comparison, results
from literature. \cite{RezRilHob-JCTC-11,gmtkn55}}
\begin{tabular}{lccccc}
  method      &H-bonds & dispersion & mixed & total & variance \\
\hline
rev-ISI        & 0.35 & 0.44 & 0.20 &  0.33 & 0.08 \\ 
ISI            & 0.37 & 0.42 & 0.19 &  0.33 & 0.09\\  
SPL            & 0.42 & 0.42 & 0.19 &  0.35 & 0.11 \\ 
LB             & 0.36 & 0.41 & 0.19 &  0.31 & 0.14  \\
\hline 
MP2            & 0.11 & 0.81 & 0.45 &  0.45 &  0.29 \\ 
SCS-MI-MP2     & 0.19 & 0.45 & 0.20 &  0.19 &  0.10 \\
SCS-CCSD       & 0.30 & 0.08 & 0.08 &  0.27 &  0.05 \\
B2PLYP         & 0.72 & 2.79 & 1.63 &  1.71 &  1.26 \\
\end{tabular}
\end{table*}

More generally, for all the SCC-ISI-like functionals
that we examined, the performance for non-covalent interactions is quite good,
being comparable or better than state-of-the art computational approaches
(see Table \ref{tab_s66maes}). Especially for dispersion and mixed complexes, all the ACMs perform 7 or 8 times better than the B2PLYP double hybrid and twice as better than MP2 (note that both of these methods have the same computational cost as the SCC-ISI-like  functionals). This is quite relevant,
considering that ISI-like functionals have not been explicitly constructed to model interaction
energies and do not employ any empirical parameter
(in contrast e.g. to the approaches in the last three lines of Table \ref{tab_s66maes}).
Notice also that, for AC-based functionals, not only is the mean absolute error low but also the variance
(last column of Tab.~\ref{tab_s66maes}). Therefore, 
these functionals can describe different types of interactions with similar accuracy.

We have shown that exchange-correlation functionals built by approximating the
adiabatic connection integrand with functions in which the input ingredients
enter in a non-linear way can be made size-consistent at no extra
computational cost. The starting idea is that size-consistency is restored once we consider fragments that are infinitely far apart, whose energy, by virtue of eq~\ref{eq:mstarACM}, we compute from the sum of quantities of individual fragments. We focused here only
on the case of non-covalent interactions, but the method is generally
applicable also to covalent systems. We also remark that, even though in this work we only consider a few ACMs functionals, the SCC based on eq~\ref{eq:mstarACM} has a more general applicability to any functional built using the adiabatic connection framework as well as to any functional depending non-linearly on size-consistent global quantities.

We have shown that our SCC provides in many cases an important correction to
the interaction energy and leads to a considerable improvement of the accuracy
of various ACMs. Thus, it is a simple and efficient way to correct one of the
main drawbacks of actual ACMs, which can now be reliably used for different
applications. This opens the quest for the development of improved ACMs. A
promising route in this direction is the construction of approximations by
interpolating energy densities along the adiabatic connection, which requires
non-local functionals for the strong-interaction limit
\cite{WagGor-PRA-14,BahZhoErn-JCP-16} and/or for the $\lambda=1$ case
.\cite{VucGor-JPCL-17}

The Supporting Information is available free of charge on the
ACS Publications website at DOI:.

\begin{acknowledgement}
This work was supported by the Netherlands Organization for Scientific Research (NWO) through an ECHO grant (717.013.004) and the European Research Council under H2020/ERC Consolidator Grant corr-DFT (Grant No. 648932).
\end{acknowledgement}

\clearpage
\onecolumn

\section{Supporting Information}

\subsection{Mathematical forms of the used interpolation models}
In this section we give mathematical forms of the used interpolation models for the AC integrand. 

\noindent {\bf Interaction-Strength Interpolation (ISI)}: \cite{SeiPerKur-PRA-00,FabGorSeiSal-JCTC-16}
\beq
\label{eq:ISI}
W_\lambda^{\rm ISI}=W_\infty[\rho]+\frac{X[\rho]}{\sqrt{1+Y[\rho]\lambda}+Z[\rho]},
\eeq
where $X$, $Y$ and $Z$ are given by:
\beq
X=\frac{x y^2}{z^2}, ~~~~~ Y=\frac{x^2 y^2}{z^4}, ~~~~~ Z=\frac{x y^2}{z^3}-1
\eeq
with $x=-2W'_0[\rho]$, $y=W'_\infty[\rho]$, and $z=W_0[\rho]-W_\infty[\rho]$. 

\noindent {\bf Revised Interaction-Strength Interpolation (rev-ISI)}:\cite{GorVigSei-JCTC-09}
\beq
\label{eq:revISI}
W_\lambda^{\rm rISI}=\frac{\partial}{\partial\lambda}~\Big(a^{\rm rISI}[\rho]\lambda\frac{b^{\rm rISI}[\rho]\lambda}{\sqrt{1+c^{\rm rISI}[\rho]\lambda}+d^{\rm rISI}[\rho]} \Big),
\eeq
with:
\begin{eqnarray}
a^{\rm rISI}[\rho]& = & W_\infty[\rho] \nonumber \\
b^{\rm rISI}[\rho] & = & -\frac{4W'_0[\rho]W'_\infty[\rho]^2}{(W_0[\rho]-W_\infty[\rho])^2} \nonumber \\
c^{\rm rISI}[\rho] & = & -\frac{8W'_0[\rho]^2W'_\infty[\rho]^2}{(W_0[\rho]-W_\infty[\rho])^4} \nonumber \\
d^{\rm rISI}[\rho] & = & -1-\frac{4W'_0[\rho]W'_\infty[\rho]^2}{(W_0[\rho]-W_\infty[\rho])^3}.e
\label{eq:revABC}
\end{eqnarray}

\noindent {\bf Seidl-Perdew-Levy (SPL)}: \cite{SeiPerLev-PRA-99,SeiGorSav-PRA-07,locpaper}
\beq
\label{eq:SPL}
W_\lambda^{\rm SPL}=a^{\rm SPL}[\rho] + \frac{b^{\rm SPL}[\rho]}{\sqrt{1 + c^{\rm SPL}[\rho] \lambda}}
\eeq
with:
\begin{eqnarray}
a^{\rm SPL}[\rho] & = & W_\infty[\rho] \nonumber \\
b^{\rm SPL}[\rho] & = & W_{0}[\rho] - W_{\infty}[\rho] \nonumber \\
c^{\rm SPL}[\rho] & = & -\frac{2 W_{0}'[\rho]}{W_{0}[\rho] - W_{\infty}[\rho]}.
\label{eq:SPLabc}
\end{eqnarray}

\noindent {\bf Liu-Burke (LB)}: \cite{LiuBur-PRA-09,locpaper}
\beq
\label{eq:LB}
W_\lambda^{\rm LB}=a^{\rm LB}[\rho] + b^{\rm LB}[\rho] \left(\frac{1}{(1 + c^{\rm LB}[\rho] \lambda)^{2}} + \frac{1}{\sqrt{1 + c^{\rm LB}[\rho] \lambda}}\right),
\eeq
with:
\begin{eqnarray}
a^{\rm LB}[\rho] & = & W_\infty[\rho] \nonumber \\
b^{\rm LB}[\rho] & = & \frac{W_{0}[\rho] - W_{\infty}[\rho]}{2} \nonumber \\
c^{\rm LB}[\rho] & = & -\frac{4 W_{0}'[\rho]}{5(W_{0}[\rho] - W_{\infty}[\rho])}
\label{eq:LBabc}
\end{eqnarray}

\subsection{Additional computational details}
The point-charge-plus-continuum (PC) functional approximations to the strong coupling limit quantities are given by:\cite{SeiPerKur-PRA-00}
\begin{eqnarray}
W_\infty[\rho] & = & \int \left[A\rho(\mathbf{r})^{4/3} + B\frac{|\nabla\rho(\mathbf{r})|^2}{\rho(\mathbf{r})^{4/3}}\right]\mathrm{d}\mathbf{r} \\
W'_\infty[\rho] & = & \int \left[C\rho(\mathbf{r})^{3/2} + D\frac{|\nabla\rho(\mathbf{r})|^2}{\rho^{7/6}(\mathbf{r})}\right]\mathrm{d}\mathbf{r}.
\label{eq:WprimePC}
\end{eqnarray}
The parameters $A=-1.451$, $B=5.317\times10^{-3}$, and $C=1.535$, are
determined by the electrostatic arguments\cite{SeiPerKur-PRA-00} and
$D=-2.8957\times10^{-2}$ has been obtained by ensuring that the given
approximation to $W'_\infty[\rho]$ is exact for the helium
atom.\cite{GorVigSei-JCTC-09} All interaction energies reported in the letter
have been corrected for the basis-set superposition error. In all calculations
(except for Kr which used an aug-cc-pV5Z basis set \cite{ccpv5z})
we used a basis set constructed adding selected $s$, $p$, $d$, and $f$ functions to
the aug-cc-pVQZ basis set \cite{ccpvqz_1,ccpvqz_2} of each element. 
The list of additional functions is reported in Table \ref{tab_basis}.

%
\begin{table}
\caption{\label{tab_basis}List of additional (Gaussian) basis functions used for each
  element.}
\begin{tabular}{lcr}
\hline\hline
Element & \multicolumn{2}{c}{Basis function}\\
\cline{2-3}
 & type $\;\;\;$ & Exponent \\
\hline
H  & $s$ &  6.17937 \\ 
   & $s$ &  0.46550 \\
   & $p$ &  3.43000 \\
   & $d$ &  4.45300 \\
He & $s$ &  19.0385 \\  
   & $s$ &   2.0880 \\
   & $p$ &  16.1040 \\
   & $p$ &   2.4980 \\
   & $d$ &  12.4980 \\
N  & $s$ &  13.8234 \\
   & $s$ &   2.1950 \\
   & $p$ &   2.1480 \\
   & $d$ &   6.7170 \\
C  & $s$ &   9.9641 \\
   & $s$ &   1.6560 \\
   & $p$ &   1.5040 \\
   & $d$ &   4.5420 \\
O  & $s$ &  18.3030 \\
   & $s$ &   2.7760 \\
   & $p$ &   2.7320 \\
   & $d$ &   8.2530 \\
Ne & $s$ &  29.0669 \\  
   & $s$ &   4.3270 \\   
   & $p$ &   4.2810 \\  
   & $d$ &  13.3170 \\  
Ar & $s$ &   1.7580 \\
   & $p$ &   2.2450 \\
   & $d$ &   4.7760 \\   
   & $f$ &   3.0582 \\
\hline\hline
\end{tabular}
\end{table}


\subsection{Results for the S66 test set}
Full results for the S66 test are reported in Tables \ref{s66all} and \ref{s66all2}

\begin{scriptsize}
\begin{table}
\caption{\label{s66all} Signed errors in kcal/mol for the S66 test for all the AC-based functionals. 
Systems 1-23 have H-bond interaction, systems 24-46 dispersion, system 47-66
mixed characters. [Continues in Table \ref{s66all2}]\\
}
\begin{tabular}{llrrrr}
\hline
num. & system   	    & rev-ISI &  ISI & 	SPL & LB \\
\hline
1 & Water-Water	&-0.064	& -0.095 &	-0.161&  -0.146 \\
2 & Water-MeOH	&-0.141	& -0.164 & 	-0.213& -0.175 \\
3 & Water-MeNH2	&-0.193	& -0.210 &	-0.245&	-0.189 \\
4 & Water-Peptide	&-0.259	& -0.294 &	-0.370&	-0.334 \\
5 & MeOH-MeOH	&-0.222	& -0.239 &	-0.276&	-0.225 \\
6 & MeOH-MeNH2	&-0.364	& -0.373 &	-0.393&	-0.302 \\
7 & MeOH-Peptide	&-0.417	& -0.442 &	-0.495&	-0.432 \\ 
8 & MeOH-Water	&-0.124	& -0.150 &	-0.204&	-0.179 \\
9 & MeNH2-MeOH	&-0.273	& -0.287 &	-0.315&	-0.261 \\
10 & MeNH2-MeNH2	&-0.376	& -0.380 &	-0.390&	-0.293 \\
11 & MeNH2-Peptide	&-0.484	& -0.487 &	-0.494&	-0.380 \\
12 & MeNH2-Water	&-0.210	& -0.223 &	-0.253&	-0.180 \\
13 & Peptide-MeOH	&-0.327	& -0.336 & 	-0.355&	-0.272 \\
14 & Peptide-MeNH2	&-0.470	& -0.466 &	-0.460&	-0.338 \\
15 & Peptide-Peptide	&-0.549	& -0.557 &	-0.575&	-0.467 \\
16 & Peptide-Water	&-0.156	& -0.179 &	-0.226&	-0.192 \\
17 & Uracil-Uracil	&-0.649	& -0.687 &	-0.767&	-0.685 \\
18 & Water-Pyridine	&-0.198	& -0.209 &	-0.234&	-0.169 \\
19 & MeOH-Pyridine	&-0.296	& -0.298 &	-0.301&	-0.206 \\
20 & AcOH-AcOH	&-0.481	& -0.556 &	-0.713&	-0.686 \\ 
21 & AcNH2-AcNH2	&-0.679	& -0.724 &	-0.819&	-0.755 \\
22 & AcOH-Uracil	&-0.550	 & -0.610 &	-0.739&	-0.693 \\ 
23 & AcNH2-Uracil	    & -0.595 & 	-0.647 &-0.756 &	-0.699 \\ 
\hline
24 & Benzene-Benzene	    & 0.268 &	0.354 &	0.531&	0.854 \\ 
25 & Pyridine-Pyridine   & 0.350 &	0.448 &	0.652&	0.998 \\ 
26 & Uracil-Uracil	    &-0.725 &	-0.617 & -0.394	& -0.002 \\
27 & Benzene-Pyridine    & 0.300 &	0.392 &	0.583 &	0.919 \\
28 & Benzene-Uracil	& -0.180 &	-0.065&	0.173 &	0.567 \\ 
29 & Pyridine-Uracil	& -0.125 &	-0.015&	0.214 &	0.594 \\
30 & Benzene-Ethene	& 0.010	 & 0.046 &	0.120 &	0.306 \\
31 & Uracil-Ethene	& -0.240 &	-0.205	& -0.134 &	0.046 \\
32 & Uracil-Ethyne	& -0.082 &	-0.056	& -0.003 & 	0.152 \\
33 & Pyridine-Ethene	& 0.019	& 0.059	& 0.143	& 0.337\\
34 &Pentane-Pentane	& -0.931 & -0.913 &	-0.876& 	-0.634 \\
35 &Neopentane-Pentane &	-0.671 & 	-0.663	& -0.649& 	-0.486\\
36 &Neopentane-Neopentane &	-0.498 &	-0.500	& -0.502 &	-0.394\\
37 &Cyclopentane-Neopentane	& -0.630 &-0.622	& -0.605 &	-0.449\\
38 &Cyclopentane-Cyclopentane & -0.736 & -0.721	&-0.690	&-0.502\\
39& Benzene-Cyclopentane & -0.265 &	-0.215	& -0.111 &	0.152\\
40& Benzene-Neopentane   & 	-0.233 & -0.202	& -0.138 &	0.061\\
41&Uracil-Pentane	     & -0.905 &	-0.844	& -0.718 &	-0.416\\
\hline
\end{tabular}
\end{table}

\begin{table}
\caption{\label{s66all2} [Continues from Table \ref{s66all}] Signed errors in kcal/mol for the S66 test for all the AC-based functionals. 
Systems 1-23 have H-bond interaction, systems 24-46 dispersion, system 47-66
mixed characters.\\
}
\begin{tabular}{llrrrr}
\hline
num. & system   	    & rev-ISI &  ISI & 	SPL & LB \\
\hline
42&Uracil-Cyclopentane  &	-0.763 & -0.707	& -0.591 &	-0.322 \\
43&Uracil-Neopentane    & -0.660 &	-0.625 &	-0.552& 	-0.348 \\
44&Ethene-Pentane	     & -0.411 &	-0.409 &	-0.406	&-0.281 \\
45&Ethyne-Pentane	     & -0.148 &	-0.144 &	-0.134	&-0.018 \\
46&Peptide-Pentane	     & -0.902 &	-0.875 &	-0.817	&-0.578 \\
\hline
47 & Benzene-Benzene	     & 0.005 &	0.039 &	0.107 &	0.287 \\
48 & Pyridine-Pyridine    & -0.022 &	0.010 &	0.077 &	0.255 \\
49 & Benzene-Pyridine     & -0.002 &	0.029 &	0.093 &	0.268 \\
50 & Benzene-Ethyne	& 0.119	& 0.125	& 0.137 &	0.238 \\
51 & Ethyne-Ethyne	& 0.037	& 0.023	& -0.005 &	0.021 \\
52 & Benzene-AcOH	& -0.192 &	-0.175 &	-0.140 &	0.002 \\
53 & Benzene-AcNH2	& -0.239 &	-0.234 &	-0.222 &	-0.106 \\
54 & Benzene-Water	& -0.091 &	-0.094 &	-0.100 &	-0.020 \\
55 & Benzene-MeOH	& -0.197 &	-0.177 &	-0.135 &	0.023 \\
56 & Benzene-MeNH2	& -0.178 &	-0.155 &	-0.108 & 	0.056 \\
57 & Benzene-Peptide	& -0.218 &	-0.175 &	-0.086 &	0.141 \\
58 & Pyridine-Pyridine & -0.299 &	-0.299 &	-0.299 &	-0.206 \\
59 & Ethyne-Water	& 0.019	 &      -0.010 &	-0.071 &	-0.071 \\
60 & Ethyne-AcOH	& -0.156 &	-0.180 &	-0.229 &	-0.177 \\
61 & Pentane-AcOH	& -0.642 &	-0.629 &	-0.603 &	-0.441 \\
62 & Pentane-AcNH2	& -0.758 &	-0.742 &	-0.709 &	-0.524 \\
63 & Benzene-AcOH	& -0.176 &	-0.141 &	-0.070 &	0.123 \\ 
64 & Peptide-Ethene	& -0.346 &	-0.345 &	-0.344 &	-0.233 \\
65 & Pyridine-Ethyne	& -0.040 &	-0.054 &	-0.086 & 	-0.047 \\
66 & MeNH2-Pyridine	 &-0.243 &	-0.220 &	-0.173 &	-0.008 \\
\hline
\end{tabular}
\end{table}
\end{scriptsize}

\bibliography{all}

\providecommand*\mcitethebibliography{\thebibliography}
\csname @ifundefined\endcsname{endmcitethebibliography}
  {\let\endmcitethebibliography\endthebibliography}{}
\begin{mcitethebibliography}{62}
\providecommand*\natexlab[1]{#1}
\providecommand*\mciteSetBstSublistMode[1]{}
\providecommand*\mciteSetBstMaxWidthForm[2]{}
\providecommand*\mciteBstWouldAddEndPuncttrue
  {\def\EndOfBibitem{\unskip.}}
\providecommand*\mciteBstWouldAddEndPunctfalse
  {\let\EndOfBibitem\relax}
\providecommand*\mciteSetBstMidEndSepPunct[3]{}
\providecommand*\mciteSetBstSublistLabelBeginEnd[3]{}
\providecommand*\EndOfBibitem{}
\mciteSetBstSublistMode{f}
\mciteSetBstMaxWidthForm{subitem}{(\alph{mcitesubitemcount})}
\mciteSetBstSublistLabelBeginEnd
  {\mcitemaxwidthsubitemform\space}
  {\relax}
  {\relax}

\bibitem[Kohn and Sham(1965)Kohn, and Sham]{KohSha-PR-65}
Kohn,~W.; Sham,~L.~J. Self-Consistent Equations Including Exchange and
  Correlation Effects. \emph{Phys. Rev.} \textbf{1965}, \emph{140}, A
  1133\relax
\mciteBstWouldAddEndPuncttrue
\mciteSetBstMidEndSepPunct{\mcitedefaultmidpunct}
{\mcitedefaultendpunct}{\mcitedefaultseppunct}\relax
\EndOfBibitem
\bibitem[Mardirossian and Head-Gordon(2017)Mardirossian, and
  Head-Gordon]{MarHea-MP-17}
Mardirossian,~N.; Head-Gordon,~M. Thirty years of density functional theory in
  computational chemistry: an overview and extensive assessment of 200 density
  functionals. \emph{Mol. Phys.} \textbf{2017}, \emph{115}, 2315--2372\relax
\mciteBstWouldAddEndPuncttrue
\mciteSetBstMidEndSepPunct{\mcitedefaultmidpunct}
{\mcitedefaultendpunct}{\mcitedefaultseppunct}\relax
\EndOfBibitem
\bibitem[Della~Sala et~al.(2016)Della~Sala, Fabiano, and Constantin]{metarev}
Della~Sala,~F.; Fabiano,~E.; Constantin,~L.~A. Kinetic-energy-density dependent
  semilocal exchange-correlation functionals. \emph{Int. J. Quantum Chem.}
  \textbf{2016}, \emph{116}, 1641--1694\relax
\mciteBstWouldAddEndPuncttrue
\mciteSetBstMidEndSepPunct{\mcitedefaultmidpunct}
{\mcitedefaultendpunct}{\mcitedefaultseppunct}\relax
\EndOfBibitem
\bibitem[Cohen et~al.(2012)Cohen, Mori-S\'anchez, and Yang]{CohMorYan-CR-12}
Cohen,~A.~J.; Mori-S\'anchez,~P.; Yang,~W. Challenges for density functional
  theory. \emph{Chem. Rev.} \textbf{2012}, \emph{112}, 289\relax
\mciteBstWouldAddEndPuncttrue
\mciteSetBstMidEndSepPunct{\mcitedefaultmidpunct}
{\mcitedefaultendpunct}{\mcitedefaultseppunct}\relax
\EndOfBibitem
\bibitem[Burke(2012)]{Bur-JCP-12}
Burke,~K. Perspective on density functional theory. \emph{J. Chem. Phys.}
  \textbf{2012}, \emph{136}, 150901\relax
\mciteBstWouldAddEndPuncttrue
\mciteSetBstMidEndSepPunct{\mcitedefaultmidpunct}
{\mcitedefaultendpunct}{\mcitedefaultseppunct}\relax
\EndOfBibitem
\bibitem[Becke(2014)]{Bec-JCP-14}
Becke,~A.~D. Perspective: Fifty years of density-functional theory in chemical
  physics. \emph{J. Chem. Phys.} \textbf{2014}, \emph{140}, 18A301\relax
\mciteBstWouldAddEndPuncttrue
\mciteSetBstMidEndSepPunct{\mcitedefaultmidpunct}
{\mcitedefaultendpunct}{\mcitedefaultseppunct}\relax
\EndOfBibitem
\bibitem[Langreth and Perdew(1975)Langreth, and Perdew]{LanPer-SSC-75}
Langreth,~D.~C.; Perdew,~J.~P. \emph{Solid State Commun.} \textbf{1975},
  \emph{{17}}, 1425\relax
\mciteBstWouldAddEndPuncttrue
\mciteSetBstMidEndSepPunct{\mcitedefaultmidpunct}
{\mcitedefaultendpunct}{\mcitedefaultseppunct}\relax
\EndOfBibitem
\bibitem[Gunnarsson and Lundqvist(1976)Gunnarsson, and
  Lundqvist]{GunLun-PRB-76}
Gunnarsson,~O.; Lundqvist,~B.~I. Exchange and correlation in atoms, molecules,
  and solids by the spin-density-functional formalism. \emph{Phys. Rev. B}
  \textbf{1976}, \emph{{13}}, 4274\relax
\mciteBstWouldAddEndPuncttrue
\mciteSetBstMidEndSepPunct{\mcitedefaultmidpunct}
{\mcitedefaultendpunct}{\mcitedefaultseppunct}\relax
\EndOfBibitem
\bibitem[Colonna and Savin(1999)Colonna, and Savin]{ColSav-JCP-99}
Colonna,~F.; Savin,~A. Correlation energies for some two- and four-electron
  systems along the adiabatic connection in density functional theory. \emph{J.
  Chem. Phys.} \textbf{1999}, \emph{{110}}, 2828\relax
\mciteBstWouldAddEndPuncttrue
\mciteSetBstMidEndSepPunct{\mcitedefaultmidpunct}
{\mcitedefaultendpunct}{\mcitedefaultseppunct}\relax
\EndOfBibitem
\bibitem[Teale et~al.(2009)Teale, Coriani, and Helgaker]{TeaCorHel-JCP-09}
Teale,~A.~M.; Coriani,~S.; Helgaker,~T. The calculation of adiabatic-connection
  curves from full configuration-interaction densities: Two-electron systems.
  \emph{J. Chem. Phys.} \textbf{2009}, \emph{{130}}, 104111\relax
\mciteBstWouldAddEndPuncttrue
\mciteSetBstMidEndSepPunct{\mcitedefaultmidpunct}
{\mcitedefaultendpunct}{\mcitedefaultseppunct}\relax
\EndOfBibitem
\bibitem[Teale et~al.(2010)Teale, Coriani, and Helgaker]{TeaCorHel-JCP-10}
Teale,~A.~M.; Coriani,~S.; Helgaker,~T. Accurate calculation and modeling of
  the adiabatic connection in density functional theory. \emph{J. Chem. Phys.}
  \textbf{2010}, \emph{{132}}, 164115\relax
\mciteBstWouldAddEndPuncttrue
\mciteSetBstMidEndSepPunct{\mcitedefaultmidpunct}
{\mcitedefaultendpunct}{\mcitedefaultseppunct}\relax
\EndOfBibitem
\bibitem[Becke(1993)]{Bec-JCP-93a}
Becke,~A.~D. A new mixing of Hartree--Fock and local density-functional
  theories. \emph{J. Chem. Phys.} \textbf{1993}, \emph{98}, 1372\relax
\mciteBstWouldAddEndPuncttrue
\mciteSetBstMidEndSepPunct{\mcitedefaultmidpunct}
{\mcitedefaultendpunct}{\mcitedefaultseppunct}\relax
\EndOfBibitem
\bibitem[Becke(1993)]{Bec-JCP-93}
Becke,~A.~D. Density-functional thermochemistry. III. The role of exact
  exchange. \emph{J. Chem. Phys.} \textbf{1993}, \emph{98}, 5648\relax
\mciteBstWouldAddEndPuncttrue
\mciteSetBstMidEndSepPunct{\mcitedefaultmidpunct}
{\mcitedefaultendpunct}{\mcitedefaultseppunct}\relax
\EndOfBibitem
\bibitem[Perdew et~al.(1996)Perdew, Ernzerhof, and Burke]{PerErnBur-JCP-96}
Perdew,~J.~P.; Ernzerhof,~M.; Burke,~K. Rationale for mixing exact exchange
  with density functional approximations. \emph{J. Chem. Phys.} \textbf{1996},
  \emph{105}, 9982--9985\relax
\mciteBstWouldAddEndPuncttrue
\mciteSetBstMidEndSepPunct{\mcitedefaultmidpunct}
{\mcitedefaultendpunct}{\mcitedefaultseppunct}\relax
\EndOfBibitem
\bibitem[Grimme(2006)]{Gri-JCP-06}
Grimme,~S. Semiempirical hybrid density functional with perturbative
  second-order correlation. \emph{J. Chem. Phys.} \textbf{2006}, \emph{{124}},
  034108\relax
\mciteBstWouldAddEndPuncttrue
\mciteSetBstMidEndSepPunct{\mcitedefaultmidpunct}
{\mcitedefaultendpunct}{\mcitedefaultseppunct}\relax
\EndOfBibitem
\bibitem[Goerigk and Grimme(2010)Goerigk, and Grimme]{LarGri-JCTC-10}
Goerigk,~L.; Grimme,~S. Efficient and Accurate Double-Hybrid-Meta-GGA Density
  Functionalsîž Evaluation with the Extended GMTKN30 Database for General
  Main Group Thermochemistry, Kinetics, and Noncovalent Interactions. \emph{J.
  Chem. Theory Comput.} \textbf{2010}, \emph{7}, 291--309\relax
\mciteBstWouldAddEndPuncttrue
\mciteSetBstMidEndSepPunct{\mcitedefaultmidpunct}
{\mcitedefaultendpunct}{\mcitedefaultseppunct}\relax
\EndOfBibitem
\bibitem[Sharkas et~al.(2011)Sharkas, Toulouse, and Savin]{ShaTouSav-JCP-11}
Sharkas,~K.; Toulouse,~J.; Savin,~A. Double-hybrid density-functional theory
  made rigorous. \emph{J. Chem. Phys.} \textbf{2011}, \emph{{134}},
  064113\relax
\mciteBstWouldAddEndPuncttrue
\mciteSetBstMidEndSepPunct{\mcitedefaultmidpunct}
{\mcitedefaultendpunct}{\mcitedefaultseppunct}\relax
\EndOfBibitem
\bibitem[Cramer and Truhlar(2009)Cramer, and Truhlar]{CraTru-PCCP-09}
Cramer,~C.~J.; Truhlar,~D.~G. Density functional theory for transition metals
  and transition metal chemistry. \emph{Phys. Chem. Chem. Phys.} \textbf{2009},
  \emph{11}, 10757\relax
\mciteBstWouldAddEndPuncttrue
\mciteSetBstMidEndSepPunct{\mcitedefaultmidpunct}
{\mcitedefaultendpunct}{\mcitedefaultseppunct}\relax
\EndOfBibitem
\bibitem[Fabiano et~al.(2009)Fabiano, Piacenza, DâAgostino, and
  Sala]{fabiano09}
Fabiano,~E.; Piacenza,~M.; DâAgostino,~S.; Sala,~F.~D. Towards an accurate
  description of the electronic properties of the biphenylthiol/gold interface:
  The role of exact exchange. \emph{The Journal of Chemical Physics}
  \textbf{2009}, \emph{131}, 234101\relax
\mciteBstWouldAddEndPuncttrue
\mciteSetBstMidEndSepPunct{\mcitedefaultmidpunct}
{\mcitedefaultendpunct}{\mcitedefaultseppunct}\relax
\EndOfBibitem
\bibitem[Corminboeuf(2014)]{corminboeuf14}
Corminboeuf,~C. Minimizing Density Functional Failures for Non-Covalent
  Interactions Beyond van der Waals Complexes. \emph{Accounts of Chemical
  Research} \textbf{2014}, \emph{47}, 3217--3224\relax
\mciteBstWouldAddEndPuncttrue
\mciteSetBstMidEndSepPunct{\mcitedefaultmidpunct}
{\mcitedefaultendpunct}{\mcitedefaultseppunct}\relax
\EndOfBibitem
\bibitem[Fabiano et~al.(2015)Fabiano, Constantin, Cortona, and
  Della~Sala]{hapbe}
Fabiano,~E.; Constantin,~L.~A.; Cortona,~P.; Della~Sala,~F. Global Hybrids from
  the Semiclassical Atom Theory Satisfying the Local Density Linear Response.
  \emph{Journal of Chemical Theory and Computation} \textbf{2015}, \emph{11},
  122--131\relax
\mciteBstWouldAddEndPuncttrue
\mciteSetBstMidEndSepPunct{\mcitedefaultmidpunct}
{\mcitedefaultendpunct}{\mcitedefaultseppunct}\relax
\EndOfBibitem
\bibitem[Ernzerhof(1996)]{Ern-CPL-96}
Ernzerhof,~M. Construction of the adiabatic connection. \emph{Chem. Phys.
  Lett.} \textbf{1996}, \emph{{263}}, 499\relax
\mciteBstWouldAddEndPuncttrue
\mciteSetBstMidEndSepPunct{\mcitedefaultmidpunct}
{\mcitedefaultendpunct}{\mcitedefaultseppunct}\relax
\EndOfBibitem
\bibitem[Seidl et~al.(1999)Seidl, Perdew, and Levy]{SeiPerLev-PRA-99}
Seidl,~M.; Perdew,~J.~P.; Levy,~M. Strictly correlated electrons in
  density-functional theory. \emph{Phys. Rev. A} \textbf{1999}, \emph{{59}},
  51\relax
\mciteBstWouldAddEndPuncttrue
\mciteSetBstMidEndSepPunct{\mcitedefaultmidpunct}
{\mcitedefaultendpunct}{\mcitedefaultseppunct}\relax
\EndOfBibitem
\bibitem[Seidl et~al.(2000)Seidl, Perdew, and Kurth]{SeiPerKur-PRL-00}
Seidl,~M.; Perdew,~J.~P.; Kurth,~S. Simulation of all-order density-functional
  perturbation theory, using the second order and the strong-correlation limit.
  \emph{Phys. Rev. Lett.} \textbf{2000}, \emph{{84}}, 5070\relax
\mciteBstWouldAddEndPuncttrue
\mciteSetBstMidEndSepPunct{\mcitedefaultmidpunct}
{\mcitedefaultendpunct}{\mcitedefaultseppunct}\relax
\EndOfBibitem
\bibitem[Mori-Sanchez et~al.(2006)Mori-Sanchez, Cohen, and
  Yang]{MorCohYan-JCP-06}
Mori-Sanchez,~P.; Cohen,~A.~J.; Yang,~W.~T. \emph{J. Chem. Phys.}
  \textbf{2006}, \emph{{125}}, 201102\relax
\mciteBstWouldAddEndPuncttrue
\mciteSetBstMidEndSepPunct{\mcitedefaultmidpunct}
{\mcitedefaultendpunct}{\mcitedefaultseppunct}\relax
\EndOfBibitem
\bibitem[Vuckovic et~al.(2016)Vuckovic, Irons, Savin, Teale, and
  Gori-Giorgi]{locpaper}
Vuckovic,~S.; Irons,~T.~J.; Savin,~A.; Teale,~A.~M.; Gori-Giorgi,~P.
  Exchange--correlation functionals via local interpolation along the adiabatic
  connection. \emph{J. Chem. Theory Comput.} \textbf{2016}, \emph{12},
  2598--2610\relax
\mciteBstWouldAddEndPuncttrue
\mciteSetBstMidEndSepPunct{\mcitedefaultmidpunct}
{\mcitedefaultendpunct}{\mcitedefaultseppunct}\relax
\EndOfBibitem
\bibitem[Cohen et~al.(2007)Cohen, Mori-S{\'a}nchez, and Yang]{CohMorYan-JCP-07}
Cohen,~A.~J.; Mori-S{\'a}nchez,~P.; Yang,~W. Assessment and formal properties
  of exchange-correlation functionals constructed from the adiabatic
  connection. \emph{J. Chem. Phys.} \textbf{2007}, \emph{127}, 034101\relax
\mciteBstWouldAddEndPuncttrue
\mciteSetBstMidEndSepPunct{\mcitedefaultmidpunct}
{\mcitedefaultendpunct}{\mcitedefaultseppunct}\relax
\EndOfBibitem
\bibitem[Gori-Giorgi et~al.(2009)Gori-Giorgi, Vignale, and
  Seidl]{GorVigSei-JCTC-09}
Gori-Giorgi,~P.; Vignale,~G.; Seidl,~M. Electronic zero-point oscillations in
  the strong-interaction limit of density functional theory. \emph{J. Chem.
  Theory Comput.} \textbf{2009}, \emph{{5}}, 743\relax
\mciteBstWouldAddEndPuncttrue
\mciteSetBstMidEndSepPunct{\mcitedefaultmidpunct}
{\mcitedefaultendpunct}{\mcitedefaultseppunct}\relax
\EndOfBibitem
\bibitem[Liu and Burke(2009)Liu, and Burke]{LiuBur-PRA-09}
Liu,~Z.~F.; Burke,~K. Adiabatic connection in the low-density limit.
  \emph{Phys. Rev. A} \textbf{2009}, \emph{{79}}, 064503\relax
\mciteBstWouldAddEndPuncttrue
\mciteSetBstMidEndSepPunct{\mcitedefaultmidpunct}
{\mcitedefaultendpunct}{\mcitedefaultseppunct}\relax
\EndOfBibitem
\bibitem[Jiang and Engel(2010)Jiang, and Engel]{JiaEng-ZPC-10}
Jiang,~H.; Engel,~E. Orbital-dependent Representation of Correlation Energy
  Functional. \emph{Zeitschrift f{\"u}r Physikalische Chemie} \textbf{2010},
  \emph{224}, 455--466\relax
\mciteBstWouldAddEndPuncttrue
\mciteSetBstMidEndSepPunct{\mcitedefaultmidpunct}
{\mcitedefaultendpunct}{\mcitedefaultseppunct}\relax
\EndOfBibitem
\bibitem[Wigner(1934)]{Wig-PR-34}
Wigner,~E.~P. On the interaction of electrons in metals. \emph{Phys. Rev.}
  \textbf{1934}, \emph{{46}}, 1002\relax
\mciteBstWouldAddEndPuncttrue
\mciteSetBstMidEndSepPunct{\mcitedefaultmidpunct}
{\mcitedefaultendpunct}{\mcitedefaultseppunct}\relax
\EndOfBibitem
\bibitem[Wigner(1938)]{Wig-TFS-38}
Wigner,~E.~P. Effects of the electron interaction on the energy levels of
  electrons in metals. \emph{Trans. Faraday Soc.} \textbf{1938}, \emph{{34}},
  678\relax
\mciteBstWouldAddEndPuncttrue
\mciteSetBstMidEndSepPunct{\mcitedefaultmidpunct}
{\mcitedefaultendpunct}{\mcitedefaultseppunct}\relax
\EndOfBibitem
\bibitem[Mirtschink et~al.(2012)Mirtschink, Seidl, and
  Gori-Giorgi]{MirSeiGor-JCTC-12}
Mirtschink,~A.; Seidl,~M.; Gori-Giorgi,~P. Energy densities in the
  strong-interaction limit of density functional theory. \emph{J. Chem. Theory
  Comput.} \textbf{2012}, \emph{8}, 3097\relax
\mciteBstWouldAddEndPuncttrue
\mciteSetBstMidEndSepPunct{\mcitedefaultmidpunct}
{\mcitedefaultendpunct}{\mcitedefaultseppunct}\relax
\EndOfBibitem
\bibitem[Zhou et~al.(2015)Zhou, Bahmann, and Ernzerhof]{ZhoBahErn-JCP-15}
Zhou,~Y.; Bahmann,~H.; Ernzerhof,~M. Construction of exchange-correlation
  functionals through interpolation between the non-interacting and the
  strong-correlation limit. \emph{J. Chem. Phys.} \textbf{2015}, \emph{143},
  124103\relax
\mciteBstWouldAddEndPuncttrue
\mciteSetBstMidEndSepPunct{\mcitedefaultmidpunct}
{\mcitedefaultendpunct}{\mcitedefaultseppunct}\relax
\EndOfBibitem
\bibitem[Bahmann et~al.(2016)Bahmann, Zhou, and Ernzerhof]{BahZhoErn-JCP-16}
Bahmann,~H.; Zhou,~Y.; Ernzerhof,~M. The shell model for the
  exchange-correlation hole in the strong-correlation limit. \emph{J. Chem.
  Phys.} \textbf{2016}, \emph{145}, 124104\relax
\mciteBstWouldAddEndPuncttrue
\mciteSetBstMidEndSepPunct{\mcitedefaultmidpunct}
{\mcitedefaultendpunct}{\mcitedefaultseppunct}\relax
\EndOfBibitem
\bibitem[Vuckovic et~al.(2017)Vuckovic, Irons, Wagner, Teale, and
  Gori-Giorgi]{VucIroWagTeaGor-PCCP-17}
Vuckovic,~S.; Irons,~T. J.~P.; Wagner,~L.~O.; Teale,~A.~M.; Gori-Giorgi,~P.
  Interpolated energy densities{,} correlation indicators and lower bounds from
  approximations to the strong coupling limit of DFT. \emph{Phys. Chem. Chem.
  Phys.} \textbf{2017}, \emph{19}, 6169--6183\relax
\mciteBstWouldAddEndPuncttrue
\mciteSetBstMidEndSepPunct{\mcitedefaultmidpunct}
{\mcitedefaultendpunct}{\mcitedefaultseppunct}\relax
\EndOfBibitem
\bibitem[Jaramillo et~al.(2003)Jaramillo, Scuseria, and
  Ernzerhof]{JarScuErn-JCP-03}
Jaramillo,~J.; Scuseria,~G.~E.; Ernzerhof,~M. Local hybrid functionals.
  \emph{J. Chem. Phys.} \textbf{2003}, \emph{118}, 1068--1073\relax
\mciteBstWouldAddEndPuncttrue
\mciteSetBstMidEndSepPunct{\mcitedefaultmidpunct}
{\mcitedefaultendpunct}{\mcitedefaultseppunct}\relax
\EndOfBibitem
\bibitem[Maier et~al.(2016)Maier, Haasler, Arbuznikov, and
  Kaupp]{MaiHasArbKau-PCCP-16}
Maier,~T.~M.; Haasler,~M.; Arbuznikov,~A.~V.; Kaupp,~M. New approaches for the
  calibration of exchange-energy densities in local hybrid functionals.
  \emph{Phys. Chem. Chem. Phys.} \textbf{2016}, \emph{18}, 21133--21144\relax
\mciteBstWouldAddEndPuncttrue
\mciteSetBstMidEndSepPunct{\mcitedefaultmidpunct}
{\mcitedefaultendpunct}{\mcitedefaultseppunct}\relax
\EndOfBibitem
\bibitem[Vuckovic et~al.(2017)Vuckovic, Levy, and
  Gori-Giorgi]{VucLevGor-JCP-17}
Vuckovic,~S.; Levy,~M.; Gori-Giorgi,~P. Augmented potential, energy densities,
  and virial relations in the weak-and strong-interaction limits of DFT.
  \emph{J. Chem. Phys} \textbf{2017}, \emph{147}, 214107\relax
\mciteBstWouldAddEndPuncttrue
\mciteSetBstMidEndSepPunct{\mcitedefaultmidpunct}
{\mcitedefaultendpunct}{\mcitedefaultseppunct}\relax
\EndOfBibitem
\bibitem[Arbuznikov and Kaupp(2007)Arbuznikov, and Kaupp]{ArbKau-CPL-07}
Arbuznikov,~A.~V.; Kaupp,~M. Local hybrid exchange-correlation functionals
  based on the dimensionless density gradient. \emph{Chem. Phys. Lett.}
  \textbf{2007}, \emph{440}, 160--168\relax
\mciteBstWouldAddEndPuncttrue
\mciteSetBstMidEndSepPunct{\mcitedefaultmidpunct}
{\mcitedefaultendpunct}{\mcitedefaultseppunct}\relax
\EndOfBibitem
\bibitem[Bahmann and Kaupp(2015)Bahmann, and Kaupp]{BahKau-JCTC-15}
Bahmann,~H.; Kaupp,~M. Efficient Self-Consistent implementation of local hybrid
  Functionals. \emph{J. Chem. Theory Comput.} \textbf{2015}, \emph{11},
  1540--1548\relax
\mciteBstWouldAddEndPuncttrue
\mciteSetBstMidEndSepPunct{\mcitedefaultmidpunct}
{\mcitedefaultendpunct}{\mcitedefaultseppunct}\relax
\EndOfBibitem
\bibitem[Gori-Giorgi and Savin(2008)Gori-Giorgi, and Savin]{GorSav-JPCS-08}
Gori-Giorgi,~P.; Savin,~A. Degeneracy and size consistency in electronic
  density functional theory. \emph{J. Phys.: Conf. Ser.} \textbf{2008},
  \emph{{117}}, 012017\relax
\mciteBstWouldAddEndPuncttrue
\mciteSetBstMidEndSepPunct{\mcitedefaultmidpunct}
{\mcitedefaultendpunct}{\mcitedefaultseppunct}\relax
\EndOfBibitem
\bibitem[Savin(2009)]{Sav-CP-09}
Savin,~A. Is size-consistency possible with density functional approximations?
  \emph{Chem. Phys.} \textbf{2009}, \emph{{356}}, 91\relax
\mciteBstWouldAddEndPuncttrue
\mciteSetBstMidEndSepPunct{\mcitedefaultmidpunct}
{\mcitedefaultendpunct}{\mcitedefaultseppunct}\relax
\EndOfBibitem
\bibitem[not()]{notedeg}
For systems with a degenerate ground-state, the presence of a system, even very
  far from another one, selects which degenerate ground state we should
  consider. In other words, each possible $M^*$ selects a different set of
  degenerate states for the various $A_i$. The exact functional should be able
  to give the same ground-state energy for all the degenerate states of the
  fragments, something that no present XC approximation is able to do. This is
  interlinked with the static correlation problem in DFT.\relax
\mciteBstWouldAddEndPunctfalse
\mciteSetBstMidEndSepPunct{\mcitedefaultmidpunct}
{}{\mcitedefaultseppunct}\relax
\EndOfBibitem
\bibitem[Seidl et~al.(2007)Seidl, Gori-Giorgi, and Savin]{SeiGorSav-PRA-07}
Seidl,~M.; Gori-Giorgi,~P.; Savin,~A. Strictly correlated electrons in
  density-functional theory: A general formulation with applications to
  spherical densities. \emph{Phys. Rev. A} \textbf{2007}, \emph{{75}},
  042511\relax
\mciteBstWouldAddEndPuncttrue
\mciteSetBstMidEndSepPunct{\mcitedefaultmidpunct}
{\mcitedefaultendpunct}{\mcitedefaultseppunct}\relax
\EndOfBibitem
\bibitem[Liu and Burke(2009)Liu, and Burke]{LiuBur-JCP-09}
Liu,~Z.~F.; Burke,~K. Adiabatic connection for strictly correlated electrons.
  \emph{J. Chem. Phys.} \textbf{2009}, \emph{{131}}, 124124\relax
\mciteBstWouldAddEndPuncttrue
\mciteSetBstMidEndSepPunct{\mcitedefaultmidpunct}
{\mcitedefaultendpunct}{\mcitedefaultseppunct}\relax
\EndOfBibitem
\bibitem[Hobza and Zahradnik(1988)Hobza, and Zahradnik]{Hob-CR-88}
Hobza,~P.; Zahradnik,~R. Intermolecular interactions between medium-sized
  systems. Nonempirical and empirical calculations of interaction energies.
  Successes and failures. \emph{Chem. Rev.} \textbf{1988}, \emph{88},
  871--897\relax
\mciteBstWouldAddEndPuncttrue
\mciteSetBstMidEndSepPunct{\mcitedefaultmidpunct}
{\mcitedefaultendpunct}{\mcitedefaultseppunct}\relax
\EndOfBibitem
\bibitem[Fabiano et~al.(2016)Fabiano, Gori-Giorgi, Seidl, and
  Della~Sala]{FabGorSeiSal-JCTC-16}
Fabiano,~E.; Gori-Giorgi,~P.; Seidl,~M.; Della~Sala,~F. Interaction-Strength
  Interpolation Method for Main-Group Chemistry: Benchmarking, Limitations, and
  Perspectives. \emph{J. Chem. Theory Comput} \textbf{2016}, \emph{12},
  4885--4896\relax
\mciteBstWouldAddEndPuncttrue
\mciteSetBstMidEndSepPunct{\mcitedefaultmidpunct}
{\mcitedefaultendpunct}{\mcitedefaultseppunct}\relax
\EndOfBibitem
\bibitem[\'Smiga and Fabiano(2017)\'Smiga, and Fabiano]{smiga17}
\'Smiga,~S.; Fabiano,~E. Approximate solution of coupled cluster equations:
  application to the coupled cluster doubles method and non-covalent
  interacting systems. \emph{Phys. Chem. Chem. Phys.} \textbf{2017}, \emph{19},
  30249--30260\relax
\mciteBstWouldAddEndPuncttrue
\mciteSetBstMidEndSepPunct{\mcitedefaultmidpunct}
{\mcitedefaultendpunct}{\mcitedefaultseppunct}\relax
\EndOfBibitem
\bibitem[Ogilvie and Wang(1993)Ogilvie, and Wang]{OgiWan-JMS-93}
Ogilvie,~J.; Wang,~F.~Y. Potential-energy functions of diatomic molecules of
  the noble gases: II. Unlike nuclear species. \emph{J. Mol. Struct.}
  \textbf{1993}, \emph{291}, 313--322\relax
\mciteBstWouldAddEndPuncttrue
\mciteSetBstMidEndSepPunct{\mcitedefaultmidpunct}
{\mcitedefaultendpunct}{\mcitedefaultseppunct}\relax
\EndOfBibitem
\bibitem[tur()]{turbo}
Available from http://www.turbomole.com (accessed Nov. 2015), TURBOMOLE, V6.3;
  TURBOMOLE GmbH: Karlsruhe, Germany, 2011\relax
\mciteBstWouldAddEndPuncttrue
\mciteSetBstMidEndSepPunct{\mcitedefaultmidpunct}
{\mcitedefaultendpunct}{\mcitedefaultseppunct}\relax
\EndOfBibitem
\bibitem[Furche et~al.(2014)Furche, Ahlrichs, H\"attig, Klopper, Sierka, and
  Weigend]{turbo_rev}
Furche,~F.; Ahlrichs,~R.; H\"attig,~C.; Klopper,~W.; Sierka,~M.; Weigend,~F.
  Turbomole. \emph{WIREs Comput. Mol. Sci.} \textbf{2014}, \emph{4}, 91\relax
\mciteBstWouldAddEndPuncttrue
\mciteSetBstMidEndSepPunct{\mcitedefaultmidpunct}
{\mcitedefaultendpunct}{\mcitedefaultseppunct}\relax
\EndOfBibitem
\bibitem[Giarrusso et~al.(2018)Giarrusso, Gori-Giorgi, Della~Sala, and
  Fabiano]{GiaGorDelFab-JCP-18}
Giarrusso,~S.; Gori-Giorgi,~P.; Della~Sala,~F.; Fabiano,~E. Assessment of
  interaction-strength interpolation formulas for gold and silver clusters.
  \emph{J. Chem. Phys.} \textbf{2018}, \emph{148}, 134106\relax
\mciteBstWouldAddEndPuncttrue
\mciteSetBstMidEndSepPunct{\mcitedefaultmidpunct}
{\mcitedefaultendpunct}{\mcitedefaultseppunct}\relax
\EndOfBibitem
\bibitem[Seidl et~al.(2000)Seidl, Perdew, and Kurth]{SeiPerKur-PRA-00}
Seidl,~M.; Perdew,~J.~P.; Kurth,~S. Density functionals for the
  strong-interaction limit. \emph{Phys. Rev. A} \textbf{2000}, \emph{{62}},
  012502\relax
\mciteBstWouldAddEndPuncttrue
\mciteSetBstMidEndSepPunct{\mcitedefaultmidpunct}
{\mcitedefaultendpunct}{\mcitedefaultseppunct}\relax
\EndOfBibitem
\bibitem[Rez{\'a}c et~al.(2011)Rez{\'a}c, Riley, and Hobza]{RezRilHob-JCTC-11}
Rez{\'a}c,~J.; Riley,~K.~E.; Hobza,~P. S66: A well-balanced database of
  benchmark interaction energies relevant to biomolecular structures.
  \emph{Journal of chemical theory and computation} \textbf{2011}, \emph{7},
  2427--2438\relax
\mciteBstWouldAddEndPuncttrue
\mciteSetBstMidEndSepPunct{\mcitedefaultmidpunct}
{\mcitedefaultendpunct}{\mcitedefaultseppunct}\relax
\EndOfBibitem
\bibitem[Goerigk et~al.(2017)Goerigk, Hansen, Bauer, Ehrlich, Najibi, and
  Grimme]{gmtkn55}
Goerigk,~L.; Hansen,~A.; Bauer,~C.; Ehrlich,~S.; Najibi,~A.; Grimme,~S. A look
  at the density functional theory zoo with the advanced GMTKN55 database for
  general main group thermochemistry{,} kinetics and noncovalent interactions.
  \emph{Phys. Chem. Chem. Phys.} \textbf{2017}, \emph{19}, 32184--32215\relax
\mciteBstWouldAddEndPuncttrue
\mciteSetBstMidEndSepPunct{\mcitedefaultmidpunct}
{\mcitedefaultendpunct}{\mcitedefaultseppunct}\relax
\EndOfBibitem
\bibitem[Wagner and Gori-Giorgi(2014)Wagner, and Gori-Giorgi]{WagGor-PRA-14}
Wagner,~L.~O.; Gori-Giorgi,~P. Electron avoidance: A nonlocal radius for strong
  correlation. \emph{Phys. Rev. A} \textbf{2014}, \emph{90}, 052512\relax
\mciteBstWouldAddEndPuncttrue
\mciteSetBstMidEndSepPunct{\mcitedefaultmidpunct}
{\mcitedefaultendpunct}{\mcitedefaultseppunct}\relax
\EndOfBibitem
\bibitem[Vuckovic and Gori-Giorgi(2017)Vuckovic, and
  Gori-Giorgi]{VucGor-JPCL-17}
Vuckovic,~S.; Gori-Giorgi,~P. Simple fully non-local density functionals for
  the electronic repulsion energy. \emph{J. Phys. Chem. Lett.} \textbf{2017},
  \emph{8}, 2799--2805\relax
\mciteBstWouldAddEndPuncttrue
\mciteSetBstMidEndSepPunct{\mcitedefaultmidpunct}
{\mcitedefaultendpunct}{\mcitedefaultseppunct}\relax
\EndOfBibitem
\bibitem[Wilson et~al.(1999)Wilson, Woon, Peterson, and Jr.]{ccpv5z}
Wilson,~A.~K.; Woon,~D.~E.; Peterson,~K.~A.; Jr.,~T. H.~D. Gaussian basis sets
  for use in correlated molecular calculations. IX. The atoms gallium through
  krypton. \emph{J. Chem. Phys.} \textbf{1999}, \emph{110}, 7667--7676\relax
\mciteBstWouldAddEndPuncttrue
\mciteSetBstMidEndSepPunct{\mcitedefaultmidpunct}
{\mcitedefaultendpunct}{\mcitedefaultseppunct}\relax
\EndOfBibitem
\bibitem[Jr.(1989)]{ccpvqz_1}
Jr.,~T. H.~D. Gaussian basis sets for use in correlated molecular calculations.
  I. The atoms boron through neon and hydrogen. \emph{J. Chem. Phys.}
  \textbf{1989}, \emph{90}, 1007--1023\relax
\mciteBstWouldAddEndPuncttrue
\mciteSetBstMidEndSepPunct{\mcitedefaultmidpunct}
{\mcitedefaultendpunct}{\mcitedefaultseppunct}\relax
\EndOfBibitem
\bibitem[Woon and Jr.(1994)Woon, and Jr.]{ccpvqz_2}
Woon,~D.~E.; Jr.,~T. H.~D. Gaussian basis sets for use in correlated molecular
  calculations. IV. Calculation of static electrical response properties.
  \emph{J. Chem. Phys.} \textbf{1994}, \emph{100}, 2975--2988\relax
\mciteBstWouldAddEndPuncttrue
\mciteSetBstMidEndSepPunct{\mcitedefaultmidpunct}
{\mcitedefaultendpunct}{\mcitedefaultseppunct}\relax
\EndOfBibitem
\end{mcitethebibliography}

\end{document}